\documentclass[sigconf]{acmart}

\AtBeginDocument{%
  }

\copyrightyear{2026}
\acmYear{2026}
\setcopyright{cc}
\setcctype{by}
\acmConference[UIST '26]{The 39th Annual ACM Symposium on User Interface Software and Technology}{November 02--05, 2026}{Detroit, MI, USA}
\acmBooktitle{The 39th Annual ACM Symposium on User Interface Software and Technology (UIST '26), November 02--05, 2026, Detroit, MI, USA}
\acmDOI{10.1145/3830398.3830680}
\acmISBN{979-8-4007-2856-3/2026/11}

\usepackage[skip=2pt]{caption}
\usepackage{graphicx} 
\usepackage{booktabs}
\usepackage{multirow} 
\begin{document}

\newcommand{\martin}[1]{{\leavevmode\color[rgb]{0, 0, 0}{#1}}}

\newcommand{\archit}[1]{\textcolor{teal}{#1}}

\newcommand{\better}[1]{{\leavevmode\color[rgb]{0, 0.6, 0.2}{#1}}}
\newcommand{\worse}[1]{{\leavevmode\color[rgb]{0.8, 0.3, 0.3}{#1}}}

\newcommand{\system}{Print\&Fold}


\title{\system{}: Printing and Folding Shape-accurate 3D Models}


\author{Archit Kumar}
\affiliation{%
  \institution{University of Washington}
  \streetaddress{Guggenheim Hall}
  \city{Seattle}
  \state{WA}
  \country{USA}}
\email{akumar57@uw.edu}

\author{Zachary Grimm}
\affiliation{%
  \institution{University of Washington}
  \streetaddress{Guggenheim Hall}
  \city{Seattle}
  \state{WA}
  \country{USA}}
\email{zgrimm@uw.edu}

\author{Mingsheng Xu}
\affiliation{%
  \institution{University of Washington}
  \streetaddress{Guggenheim Hall}
  \city{Seattle}
  \state{WA}
  \country{USA}}
\email{mingsx@uw.edu}

\author{Shlok Rathi}
\affiliation{%
  \institution{University of Washington}
  \streetaddress{Guggenheim Hall}
  \city{Seattle}
  \state{WA}
  \country{USA}}
\email{srathiuw@uw.edu}

\author{Martin Nisser}
\affiliation{%
  \institution{University of Washington}
  \streetaddress{Guggenheim Hall}
  \city{Seattle}
  \state{WA}
  \country{USA}}
\email{nisser@uw.edu}

\renewcommand{\shortauthors}{Kumar et al.}

\begin{abstract}

This paper introduces \system{}, a tool to allow FDM 3D printing of complex models with less time and material while preserving shape accuracy. Key to this work is a folding algorithm that planarizes foldable faces \textit{internal} to the 3D model. While folding techniques typically discretize a target model's surface, thereby fabricating low fidelity counterparts, our method preserves the surface features in the physical print. Our design tool allows users to unfold 3D models to be FDM-printed flat before manually folding these into their target shapes. We showcase a variety of applications and evaluate the material and time savings across a range of 3D models. 

\end{abstract}


\begin{teaserfigure}
  \includegraphics[width=\textwidth]{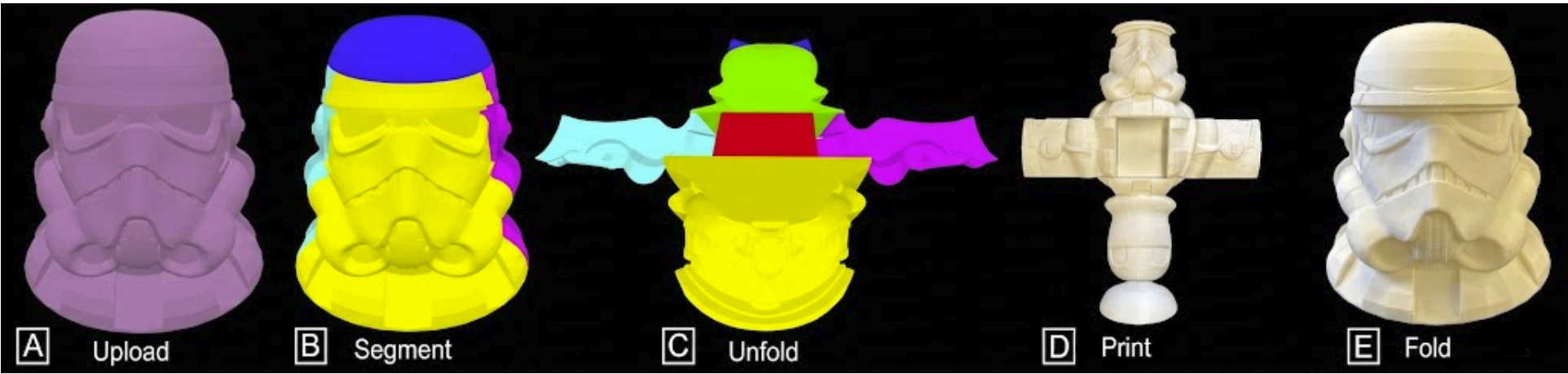}
  \caption{\system{} Pipeline. Users (A) upload a 3D model and (B) segment it into foldable panels which are (C) unfolded. The unfolded geometry is (D) FDM 3D printed flat and (E) folded into its 3D configuration without loss of surface resolution.}
  \Description{Pipeline.}
  \label{fig:fig1}
\end{teaserfigure}


\maketitle

\section{Introduction}

In personal fabrication, the goal of rapid prototyping is to construct an object inexpensively, rapidly, and  accurately. Efforts in HCI to improve the speed and material costs of rapid prototyping span fabrication modalities that include 3D printing wireframes~\cite{mueller2014wireprint}, laser cutting foldable panels~\cite{abdullah2022hingecore}, interactive shape displays~\cite{suzuki2018dynablock}, and wirebending metals~\cite{liu2017wirefab}. However, fabricating objects rapidly and inexpensively often comes at the \textit{expense} of shape accuracy, resulting in low fidelity prototypes of complex digital 3D models.  

Of particular interest is to reduce material costs and time\textemdash while preserving accuracy\textemdash for 3D printing, specifically Fused Deposition Modelling (FDM), as this remains the most inexpensive and popular form of personal fabrication machine~\cite{roudny2022thermal,sargent20193d}. Reducing material lowers not just financial costs borne by the user, but environmental costs at large, as the most widely used FDM materials, PLA and ABS, suffer from significant recycling challenges~\cite{rivera2023designing}. Reducing material consumption typically leads directly to reducing fabrication time. Shorter rapid prototyping times have been shown to improve design quality, iteration, and exploratory breadth~\cite{lim2008anatomy}, enabling designers to evaluate more physical
concepts in less time.

Folding methods are a compelling solution to saving time and material, particularly when integrated with 3D printing. By printing 3D models in flat configurations and folding them into 3D, savings can accrue in material and time by avoiding infill. However, existing techniques face three main limitations. (1) Material and time savings are typically paid for by a deterioration in shape accuracy, as folding methods discretize the model surface into foldable faces. (2) No folding algorithm is known that can unfold arbitarily complex 3D meshes for fabrication, without approximating surfaces as polyhedral manifolds for developability. (3) Existing methods often require shape memory materials or manual intervention during printing, and are not unfoldable. 

In this paper, we address these limitations to allow using FDM 3D printing to fabricate reversibly foldable complex objects using less time and material, while preserving the shape accuracy of the part. \system{} enables this through a fundamentally new strategy: rather than unfolding the model's \textit{surface}, it identifies the maximum-volume rectangular prism inscribable
within the model and unfolds the model about the edges
of this \textit{interior} core. Material and time savings accrue because this hollow core replaces infill material. Because the folded panels are rigid sub-regions of the original mesh---rotated, not resampled---every surface feature is preserved exactly in the physical print, with no geometric modification. Our design tool allows users to upload a 3D model, unfold it, then export a 3D-printable file which is printed flat and manually folded. During export, the non-zero thickness meshes are thickened, living hinges are applied to folded edges, edges are chamferred to ensure correct fold angles, and peg-and-hole connectors are applied to select edges to mechanically secure the folded geometry. 
\martin{Our contributions are:
\begin{enumerate}
    \item An inscribed-box folding algorithm that unfolds complex 3D models around a hollow core, preserving shape accuracy.
    \item A tool that unfolds 3D models and automatically generates hinges and connectors to secure folded objects.
    \item An evaluation of time and material savings using our method.  
\end{enumerate}}
\martin{For the specific case of convex polyhedra, existing folding algorithms already generate volumetrically optimal material-saving cores. To supplement our results, we integrate this into our tool and evaluate the time and material savings accrued from combining this algorithm with our hinge- and connector-generation pipeline.}




\section{Related work}

Enabling users to reduce both time and material consumption in personal fabrication and rapid prototyping are key research topics in HCI. We discuss how related works address this across fabrication paradigms, then discuss folding techniques in particular.

\subsection{Saving Material and Time in Fabrication}


\textit{InFORM}~\cite{follmer2013inform} used shape displays to rapidly construct 2.5D surfaces, which \textit{Dynablock}~\cite{suzuki2018dynablock} augmented with detachment mechanisms to allow graspable 3D objects.
\textit{ProtoMold}~\cite{yamaoka2017protomold} combined shape displays with vacuum forming to rapidly mold 2.5D objects. In turn, \textit{FormFab}~\cite{FormFab} leveraged vacuum forming with a robotically controlled heat gun to mold objects in interactive time frames.

Combing disparate techniques, \textit{Stackmold}~\cite{valkeneers2019stackmold} streamlines fabrication of multi-material objects with embedded electronics by interleaving manual molding and laser cutting steps. \textit{Laserfactory}~\cite{nisser2021laserfactory} automates several manufacturing processes in one system to produce functional devices such as a quadrotor in minutes.
Exploring wirebending, \textit{Wirebend-kit}~\cite{faruqi2025wirebend} demonstrated how to wirebend custom metal wireframes to rapidly prototype low-fidelity objects. \textit{WireFab}~\cite{liu2017wirefab} increased prototyping speed by partitioning objects into low fidelity wirebent skeletons and high-fidelity 3D prints.
To accelerate prototyping of electronic products, CurveBoards~\cite{zhu2020curveboards} 3D print models as custom 3D breadboards filled with conductive silicone, allowing repositioning components on the 3D form. 

Focusing on FDM 3D printing alone, \textit{Scrappy}~\cite{wall2021scrappy} allows users to replace infill material with scrap to reduce material consumption during FDM 3D printing. \textit{WirePrint}~\cite{mueller2014wireprint} and \textit{On-The-Fly-Print}~\cite{peng2016fly} create wireframe objects using FDM 3D printing to rapidly physicalize low fidelity previews. \martin{\textit{Zip-up Print}~\cite{yamamoto2026zip} and \textit{Touch-n-Curl}~\cite{pan2025touch} print and assemble zipper-like structures to fabricate lower-fidelity artifacts using less material and time.} 
Other approaches demonstrate low-fidelity rapid prototyping by substituting bulk 3D-printed features with Lego bricks (\textit{Fabrickation}~\cite{mueller2014fabrickation}) or laser-cut sheets (\textit{Platener}~\cite{beyer2015platener}, \textit{Cofifab}~\cite{song2016cofifab}). Leveraging the speed of laser cutting over FDM printing, \textit{Roadkill}~\cite{abdullah2021roadkill} further accelerates the assembly of laser-cut parts by embedding instructions into their cutting plans. 

While these methods achieve significant speed and material improvements, all barring \textit{Scrappy}~\cite{wall2021scrappy} accomplish this using low fidelity fabrication that approximate the target model's shape.


\begin{figure*}[h]
  \centering
  \includegraphics[width=0.99\linewidth]
{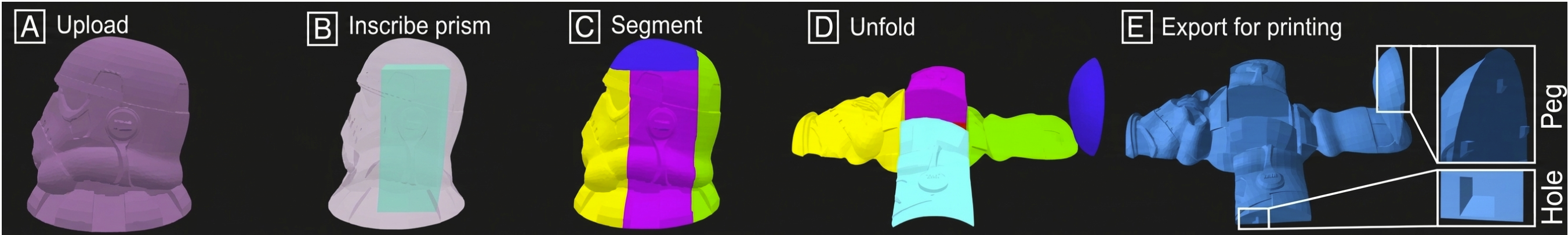}
  \caption{Complex model pipeline. Users (A) upload a model, (B) inscribe a hollow core, and (C) segment the model into parts that are (D) unfolded. The segments are (E) exported for printing with living hinges and interlocking peg-and-hole connectors.}
  \vspace{-0.5em}
  \Description{UI}
  \label{fig:UI}
\end{figure*}

\subsection{Folding in Personal Fabrication}

Seminal work on folding algorithms have demonstrated folding any polyhedral surface from paper (\textit{Origamizer}~\cite{demaine2017origamizer}), and folding patterns for several geometries from thick materials~\cite{tachi2011rigid, ku2016folding}. A core theme in HCI research centers on developing folding methods to support personal fabrication and rapid prototyping. 
For example, \textit{FabricFaces}~\cite{wagner2023fabricfaces} 3D print frames onto textiles that are folded into target shapes. \textit{Crane}~\cite{suto2023crane} allows origami design of products while considering constraints across disparate fabrication processes. 


Leveraging the speed of laser cutting, \textit{HingeCore}~\cite{abdullah2022hingecore} laser-cut foamcore with material hinges to rapidly fold orthogonal polyhedra. \textit{Flaticulation}~\cite{fang2022flaticulation} laser-cut T-patterned joints that fold over a continuous range. \textit{PullupStructs}~\cite{niu2023pullupstructs} assign fold angles via a pull-up net by routing string through laser-cut faces, allowing the structured to be folded by pulling on a single actuated string.

Researchers have also investigated self-folding, particularly for folding robots~\cite{Gaolin2026}, typically by embedding material actuators that contract across hinge lines given an environmental stimulus such as heat~\cite{felton2014method,nisser2016feedback}. HCI researchers have leveraged 3D-printable shape memory polymers to save material and time in personal fabrication. \textit{Inkjet 4D print}~\cite{narumi2023inkjet} inkjet-print onto a heat-shrinkable base which is uniformly heated to fold into 3D. \textit{OriStitch}~\cite{chang2025oristitch} combine laser cutting and embroidery with heat-shrinking thread to fold 3D textiles.
Researchers have further developed techniques to allow widely available FDM printers to print shape memory polymers as flat sheets that fold into   
approximate 3D models with curves
(\textit{Thermorph}~\cite{an2018thermorph}), large non-developable structures (\textit{4DMesh}~\cite{wang20184dmesh}), continuous double-curvature surfaces (\textit{Geodesy}~\cite{gu2019geodesy}), and 3D shapes consisting of linear elements (\textit{A-line}~\cite{wang2019line}).

Folding-related techniques have also been developed to enable time and material savings in fabricating limited 3D models with smooth surfaces.
\textit{Pop-up print}~\cite{noma2020pop} polyjet-print certain 3D models with these savings, but is constrained to 3D models that are tapered along a straight axis and have surfaces that are locally developable at the hinge.
Another strategy relies on inflatables, through custom-molding silicone balloons (\textit{Skouras et al.}~\cite{balloons}) or polyjet-printing an elastomeric shell (\textit{Blow-up Print}~\cite{matsuura2022blow}) that can be inflated to a desired shape. However, inflated objects suffer from shape inaccuracies, loss of spatial resolution, and lack of mechanical stiffness. 


In summary, folding methods suffer from several limitations, most commonly the discretization of the target surface, yielding low-polygon approximations of the desired 3D shape by reducing the physical object's surface resolution. There is no method, barring \textit{Scrappy}~\cite{wall2021scrappy}, that uses FDM printing alone to achieve full-resolution prints of complex shapes with less time and material. Unlike \textit{Scrappy} however, \system{} does not require manual intervention during printing, searching/scanning admissable objects for insertion, or printing on challenging materials and surfaces, and also enables unfolding of objects for transportation or storage.

\vspace{-0.5em}
\section{Overview of \system{}}

\system{} is a computational design tool that allows users to print and fold 3D objects with material and time savings (Figure \ref{fig:fig1}).
The design tool is a browser-based application built on Three.js
that translates a 3D model into a foldable, flat-printable counterpart in four steps, outlined below. 



\textbf{Step 1 --- Upload.} The user uploads a 3D model (.obj) by pressing \textit{Upload}. This displays the mesh in the 3D viewport, and can be zoomed and rotated for inspection using the scroller and arrowkeys. 



\textbf{Step 2 --- Segment and Unfold.}
Pressing \textit{Segment} triggers one of two decomposition pipelines that segments the model into foldable faces. The key idea behind our segmentation algorithms is to carve out the largest possible hollow internal to the 3D model---saving as much material and printing time as possible---before finding an admissible fold pattern for the remaining material. For complex 3D models, our custom inscribed-box segmentation strategy (Sec~\ref{sec:organic}) decomposes the model into 6 foldable panels around a hollow core (Figure \ref{fig:UI}-B,C). For the special case of convex polyhedra, we use a modified spanning tree approach~\cite{takahashi2011optimized,bhargava2025mesh} which decomposes the model into as many panels as there are faces (Sec~\ref{sec:polyhedral}). Both methods preserve the external shape of the model. With segmentation complete, pressing \textit{Unfold} triggers 
a real-time folding animation in the viewport. An animation
speed slider lets users inspect the folding/unfolding sequence at any tempo.

\textbf{Step 3 --- Export.}
Pressing \textit{Export} triggers a geometry processing pipeline that prepares the model for FDM printing. First, a living hinge is generated at every folding edge. Next, edges are chamferred if required for faces to fold or meet at the correct dihedral angle. Then, peg-and-hole connectors are generated to mechanicaly secure the model once folded. This processed geometry is finally saved as an .STL file for printing in its unfolded (flattened) state. 

\textbf{Step 4 --- Print \& fold.} The model is FDM-printed as one connected flat sheet and is assembled without adhesives: The user folds the print along hinge lines until they reach their mechanical stops, and connects peg-and-hole connectors to secure the final shape.



\section{\system{} Technical Pipeline}

\subsection{Complex 3D Meshes}
\label{sec:organic}

\textbf{Overview:} We wish to convert a triangular mesh $\mathcal{M}$, representing an organic 3D shape, into a foldable flat net $\mathcal{N}$, whose panel exteriors are geometrically identical to the corresponding surfaces of $\mathcal{M}$. For complex meshes, the tool runs our inscribed-box segmentation (Figure \ref{fig:UI}). This first computes the maximum volume rectangular prism that can be inscribed in the mesh. This volume is subtracted from the mesh to form the largest possible box-shaped hollow, resulting in material and time savings. The surrounding surface is then partitioned into six color-coded panels, where each panel contains one flat face coinciding with a face of the box-shaped hollow. When the model is unfolded, the flat faces will become co-planar, and are printed directly onto the buildplate. Our algorithm for partitioning panels ensures they can be folded back into the 3D shape without intersection. \martin{However, complex models are limited to those containing at least one flat surface to ensure co-planarity of unfolded faces (Figure~\ref{fig:flat-face})}; one of the hollow's faces must be inscribed in this surface, thus its size will affect the size of the hollow.



\textbf{Step 1 --- Maximum inscribed rectangular prism:}
We seek the maximum-volume axis-aligned rectangular prism $\mathcal{B}$ contained in $\mathcal{M}$, maximizing the volume efficiency ratio $\eta = V_{\mathcal{B}} / V_{\mathcal{M}}$. A rectangular prism is chosen because its six faces span three orthogonal axis pairs, partitioning $\mathbb{R}^3$ into half-spaces that allow every surface triangle of $\mathcal{M}$ to map to exactly one face by axis-dominance for overlap-free unfolding. \martin{We begin by detecting the largest planar face by clustering co-planar triangles in the mesh, then rotate the model to align it with this face.}
Finding the largest inscribed convex body in a non-convex polytope is related to the open Convex Skull Problem~\cite{crombez2018peeling}, so 
next 
we adopt a two-phase heuristic. First, we sample $k{=}64$ horizontal cross-sections of $\mathcal{M}$ at uniformly spaced heights and find the largest axis-aligned rectangle inscribed in each cross-sectional polygon, extending it vertically to produce $k$ candidate boxes. We then select the candidate with the highest $\eta$ and refine its six face positions with a gradient-free coordinate-wise search, clamping each face to remain inside $\mathcal{M}$.

\begin{figure*}[t]
  \centering
  \includegraphics[width=0.99\linewidth]
{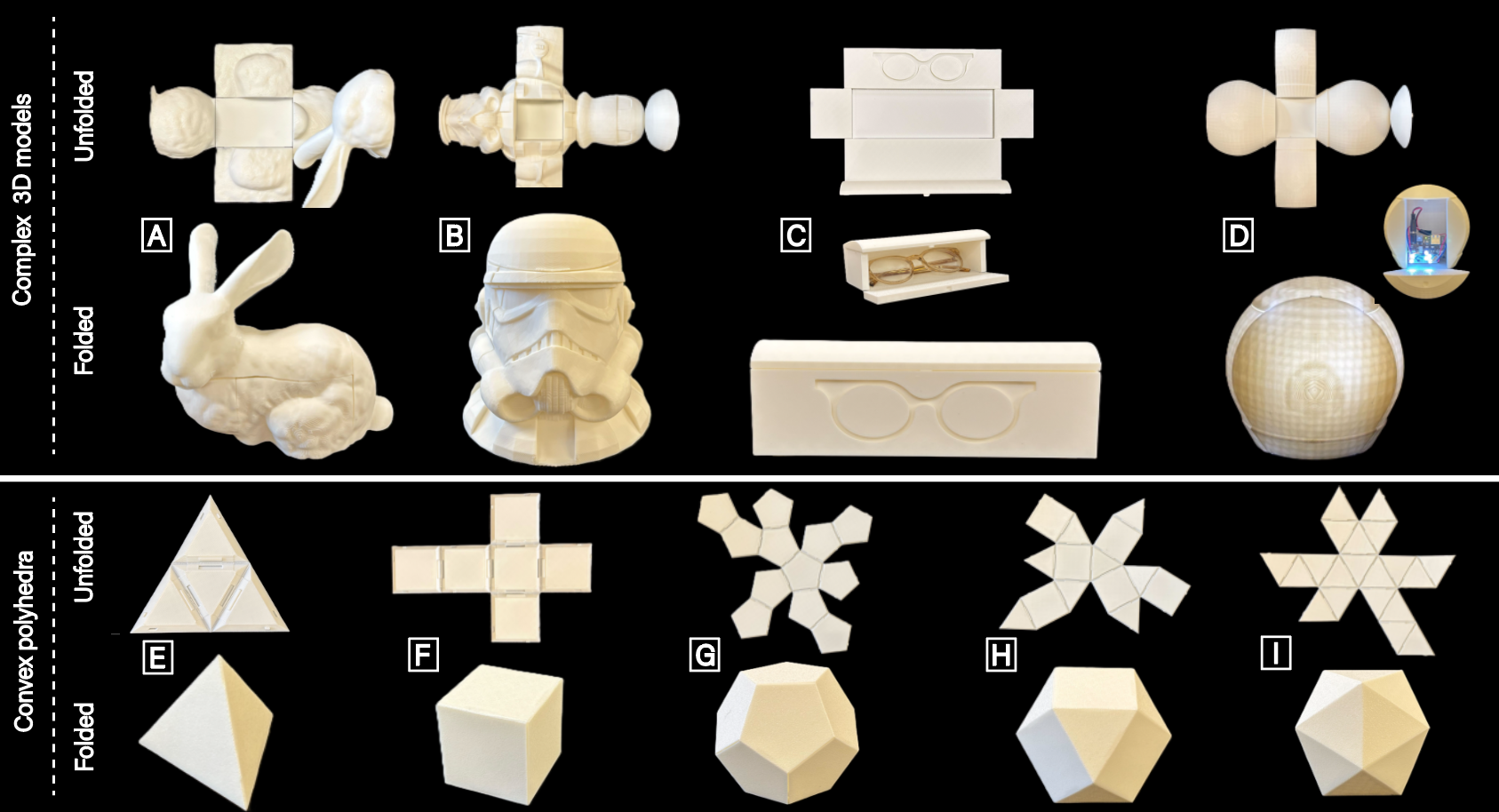}
  \caption{Unfolded and folded 3D prints. Above---Complex models: (A) Stanford bunny, (B) Stormtrooper, (C) Glasses case (with partial folding revealing glasses), (D) Lightbulb (with partial folding revealing enclosed electronics). Below---Convex polyhedra (Platonic \& Archimedean solids): (E) Tetrahedron, (F) Cube, (G) Dodecahedron, (H) Cuboctahedron, (I) Icosahedron.}
  \vspace{-0.5em}
  \Description{print}
  \label{fig:print}
\end{figure*}

\textbf{Step 2 --- Surface segmentation:}
Given $\mathcal{B}$, we partition the triangles of $\mathcal{M}$ into 6
panels $(P_i)_{i=1}^6$ corresponding to the 6 faces of $\mathcal{B}$.
Naive nearest-face assignment by centroid distance produces ragged
boundaries at corners where two faces of $\mathcal{B}$ are equidistant.
We instead use a system of clipping planes: the six outward face planes of $\mathcal{B}$ are augmented with eight diagonal bisector planes (one for each pair of adjacent box faces), forming a polyhedral classification region for each panel. Each triangle is assigned to the panel whose classification region contains the triangle's centroid.
This produces six contiguous, non-overlapping panel regions whose union exactly covers $\mathcal{M}$, with clean, predictable seam lines at the diagonal bisectors of adjacent faces.

\textbf{Step 3 --- Panel planarization:}
Let $\hat{n}_i$ denote the outward normal of face $i$ of $\mathcal{B}$,
and let $h_i$ be the corresponding face plane. Hinge edge $\ell_i$
between panel $P_i$ and the core $\mathcal{B}$ is the intersection of
$h_i$ with the mesh $\mathcal{M}$, computed by clipping each triangle
of $P_i$ against $h_i$. Each panel $P_i$ is then rotated about $\ell_i$
by the dihedral angle between its mean normal and plane $h_i$ until coplanar with $h_i$.
At the boundary of $P_i$, triangles that straddle the plane $h_i$ are clipped: the sub-triangle on the outer side of $h_i$ is retained, and new vertices are inserted by exact linear interpolation along triangle edges at their intersection with $h_i$. 


\textbf{Step 4 --- Prepare for 3D printing:}
A uniform shell thickness $t$ (default $t=0.8\,\text{mm}$, user-configurable) is extruded inward from the outer surface of each panel to produce a watertight printable solid. 
Living hinges of width $w_h = 0.2\,\text{mm}$ and thickness $t_h = 0.4\,\text{mm}$ are printed monolithically with the surrounding shell to produce a narrow elastic zone that acts as a compliant fold axis. 3 peg-and-hole connectors are generated to secure the folded model. Pegs are 5x2.5mm rectangular cross-sections of depth 5mm with one $45^\circ$ chamferred edge. The chamfer allows printing pegs onto vertical panel faces without support, and corresponding holes (toleranced to $+0.2\,\text{mm}$) are generated on mating sidewalls (Figure \ref{fig:UI}-E).




\subsection{Convex Polyhedra}
\label{sec:polyhedral}

\textbf{Overview}: For the special case of convex 3D models with planar faces---such as the Platonic and Archimedean solids (Figure \ref{fig:print}, below)---our inscribed-box decomposition method may not be optimal. For this class of models, a developable net can be found via spanning tree~\cite{takahashi2011optimized,bhargava2025mesh}, and each face can be printed without surface approximation. For convex polyhedra, this leads to the maximum volume internal hollow. We adapt this segmentation method to work with a modified hinge- and connector-generation pipeline within our tool under a unified pipeline. 
\martin{Our tool uses Quickhull, a convex hull algorithm, to check 3D models for convexity and update the default segmentation strategy in the UI accordingly. The implementation for unfolding and printing convex polyhedra is detailed in Appendix \ref{Appendix:convex}.}

\vspace{-1em}
\section{Evaluation}

\textbf{Evaluated models.}
We selected 9 models (Figure \ref{fig:print}) to evaluate our system. \martin{For complex models, we selected 4 models with favorable $\eta$ to demonstrate where our material/time savings might accrue most:} a \textit{Stormtrooper helmet} (compound-curved surface with detailed features demonstrating shape preservation); a \textit{glasses case} (demonstrating functional hollow objects that can be flat-packed for travel); a \textit{lightbulb} (demonstrating hollow enclosures for removable electronics); and the \textit{Stanford Bunny} (benchmarking against related work). For convex polyhedra, we fabricated a set of 5 Platonic and Archimedean solids---tetrahedron, cube, dodecahedron, cuboctahedron, icosahedron---normalised to an $80 \times 80 \times 80\,\text{mm}$ bounding box. All models were printed in PLA on Bambu A1/X1C FDM printers at $0.2\,\text{mm}$ layer height and $20\%$ gyroid infill on panel faces.

\begin{table}[t]
\centering
\setlength{\tabcolsep}{2pt} 
\caption{Mass \& time savings using \system{}}
\label{table1}
\noindent
\begin{tabular*}{\columnwidth}{@{\extracolsep{\fill}} l c@{\hspace{1pt}}c c@{\hspace{1pt}}c}
\toprule
\multirow{2}{*}{Model \martin{[$\eta$]} } 
& \multicolumn{2}{c}{Nominal} 
& \multicolumn{2}{c}{Print\&Fold [Gain]} \\
\cmidrule(lr){2-3} \cmidrule(lr){4-5}
 & Mass (g) & Time (m) & Mass (g) & Time (m) \\
\midrule

Stanford Bunny \martin{[0.48]} & 142 & 234 & 81  \better{[\(-\)43\%]} & 179  
  \better{[\(-\)24\%]} \\
  Stormtrooper \martin{[0.51]}  & 279 & 358 & 154 \better{[\(-\)45\%]} &
   261 \better{[\(-\)27\%]} \\
  Glasses case \martin{[0.79]}   & 125 & 129 & 82  \better{[\(-\)35\%]} & 111  
  \better{[\(-\)14\%]} \\
  Lightbulb  \martin{[0.46]}    & 397 & 346 & 233 \better{[\(-\)41\%]} & 315  
  \better{[\(-\)9\%]}  \\
  
\midrule 

Tetrahedron   & 39 & 54  & 39 [0\%] & 74
    \worse{[+37\%]} \\
  Cube          & 104 & 100 & 86 \better{[\(-\)17\%]} & 128    
  \worse{[+28\%]} \\
  Dodecahedron  & 71 & 65  & 51 \better{[\(-\)28\%]} & 85      
  \worse{[+31\%]} \\
  Cuboctahedron & 77 & 82  & 59 \better{[\(-\)23\%]} & 103     
  \worse{[+26\%]} \\
  Icosahedron   & 64 & 74  & 50 \better{[\(-\)22\%]} & 102     
  \worse{[+38\%]} \\
  
\bottomrule
\end{tabular*}
\end{table}

\textbf{Material \& time savings}: Material and time savings for each model with respect to baseline printing are shown in Table \ref{table1}. \martin{Note the inscribed-box volume-efficiency ratio $\eta = V_{\mathcal{B}}/V_{\mathcal{M}}$ applies only to the complex-model pipeline (convex polyhedra use spanning-tree unfolding)}. To establish a fair comparison, the total number of wall (2), top (4) and bottom (4) loops are consistent across conditions. For complex models, our novel \textit{inscribed-box segmentation strategy} achieves material and time savings for every model; average savings of 41\% and 19\% respectively, and best-case savings of 45\% and 27\% for the Stormtrooper. In comparison, \textit{Scrappy}~\cite{wall2021scrappy} reports average material and time savings of 29.4\% and 26.4\% respectively for their (different) objects. Benchmarking using the Stanford bunny, Pop-up print~\cite{noma2020pop} cite 5\% and 25\% material and time savings. \martin{In comparison, we yield 39\% and 18\% savings for an estimated equivalent-sized (70mm length) bunny or 43\% and 24\% savings for a 124mm length bunny, respectively.} The dominant cause of our savings is the elimination of infill material within the inscribed core. A second cause is the elimination of support material for surface features due to flat-printing: the bunny and stormtrooper require support for baseline printing, while only the bunny does in \system{}---and the support structure is different due to the rotated geometry. 
For convex polyhedra, modifying the spanning tree approach for our fabrication pipeline achieves average material savings of 18\% but time increases of 32\%. The primary cause of this tension is that while an optimal hollow shape is used to reduce infill material, the added surfaces (chamferred edges and connectors) are printed significantly slower than infill---up to 25x slower on Bambu's default settings. While nominal volume-to-area ratio increases monotonically with polyhedra faces to allow more infill removal, the additional surfaces offset some mass savings. 

\martin{To illustrate how savings scale with input model size, we re-sliced one complex model (Stanford bunny) and one convex model (cube) across three sizes (Table~\ref{tab:scaling}). As the external shell grows approximately with surface area (length$^2$), the hinges grow approximately with length, and connectors stay a fixed size, but the internal hollow grows proportionately to volume (length$^3$), larger models save proportionally more. The cube's mass savings rise from 8\% to 17\% (with its time penalty easing from +41\% to +28\%), and the Stanford bunny's mass savings rise from 38\% to 43\% (with time savings rising from 16\% to 24\%). This confirms \system{} favors larger models and matches our deduction that additional surfaces offset savings.}

\begin{table}[t]
    \centering
    \caption{\martin{Effect of complex and convex model sizes on mass and time savings. Size for the Stanford bunny is reported as the length of its major axis; for the cube, as its edge length.}}
    \label{tab:scaling}
    \begin{tabular}{llcc}
    \toprule
    Model & Length & Mass [gain] & Time [gain] \\
    \midrule
    Cube   & 40\,mm  & \better{$-$8\%}  & \worse{+41\%} \\
                    & 60\,mm  & \better{$-$14\%} & \worse{+32\%} \\
                    & 80\,mm  & \better{$-$17\%} & \worse{+28\%} \\
    \midrule
    Stanford bunny & 60\,mm  & \better{$-$38\%} & \better{$-$16\%} \\
                    & 90\,mm  & \better{$-$41\%} & \better{$-$21\%} \\
                    & 120\,mm & \better{$-$43\%} & \better{$-$24\%} \\
    \bottomrule
    \end{tabular}
  \end{table}


\textbf{Shape accuracy}: The shape accuracy of folded structures is typically a function of (1) the degree of surface discretization (the number of faces used to approximate a continuous mesh), and (2) the fold angle error between adjacent faces. For example, Thermorph~\cite{an2018thermorph} approximates the Stanford Bunny using 26 surface faces; Inkjet 4D Print~\cite{narumi2023inkjet} uses 635, attributing their accuracy gain directly to the higher face count. However \system{} is architecturally exempt from error class (1) as no surface discretization occurs. For our complex model pipeline, panels are rigid sub-regions of the input mesh, not an approximate reconstruction; for convex polyhedra, surfaces are already discretized and are not modified. In both cases, the printed outer surface is geometrically identical to the input model at printer resolution. An important exception occurs at seam contours, where adjacent panel edges meet and can be seen visually. Our system is nominally affected by error class (2) due to fold angle error at hinge lines, but our peg-and-hole connectors at leaf faces bound the global effect of odometrically accumulating angle errors below $1^\circ$ resolvable by protractor.





\textbf{Foldability \& repeatability}: We folded and unfolded a polyhedron (cube) and complex model (Stanford bunny) 20x each without observing deterioration in the mate connectors to lock folded models. We conducted a fatigue test by cyclically folding and un-folding a hinge on each model 100x times to $180^\circ$ without tearing.




\section{Discussion}



\system{} demonstrates FDM printing of shape-accurate geometries with material and time savings for complex 3D models using our inscribed prism algorithm, and with mass savings for convex polyhedra using a modified spanning tree approach. It accomplishes this with three key benefits: (1) \textit{Usability:} objects can be fabricated on inexpensive single-nozzle FDM 3D printers using PLA without intervention, and can be repeatedly folded for transport or insertion of parts or electronics into hollows. (2) \textit{Accuracy}: folded objects are shape-accurate, preserving high-resolution model features without discretization. (3) \textit{Simplicity:} transformation between flat-printed and target 3D shapes can be performed manually without post-processing. Unlike closely related work, our method preserves the model's surface features without discretization~\cite{an2018thermorph}, does not require manual intervention~\cite{wall2021scrappy}, is not limited to tapered geometries~\cite{noma2020pop}, and works with inexpensive FDM printers ~\cite{narumi2023inkjet}.







\textbf{Limitations:}
For folding complex models, the primary limitation of \system{} is that material and time savings scale with the volume efficiency ratio $\eta$: it works best for 3D models that can be inscribed with a proportionately large rectangular prism. Thin-shelled or topologically complex models, including those exhibiting concavity or strongly varying cross-sections, are expected to yield modest savings. \martin{Figure \ref{fig:Benchy} illustrates our pipeline on the popular Benchy model which exhibits a modest volume efficiency ratio of $\eta = 0.08$.}
A natural extension is to explore replacing the inscribed rectangular prism with the maximum inscribed convex polyhedron, generalizing to arbitrary convex cores and increasing $\eta$ for non-box-like models---related to the open Convex Skull Problem~\cite{crombez2018peeling}. \martin{Another limitation is the requirement of input 3D models to exhibit at least one planar face, however we observed that a variety of models satisfy this constraint to rest stably on a surface, and our demonstrations illustrate the diversity of such existing shapes.}

For folding convex polyhedra, material savings increased for all models (except tetrahedron; unchanged) but printing time deteriorated. While the economic and sustainability advantages of material savings may be sufficient, prototyping time is slowed for these objects at small scales. In addition, while our inscribed box method requires a fixed number (5) of manual folds for highly complex shapes, the number of manual folds grows with edge number for polyhedra. While our complex model pipeline may both be more efficacious and of greater utility in personal fabrication, accelerating our convex polyhedra pipeline by obviating face chamferring would complement important existing surface discretization methods.  

Finally, \system{} requires flattening 3D models onto a printbed area. Future work can explore how to utilize this compact stowage state to facilitate remote assembly~\cite{nisser2022selective,nisser2022electrovoxel} and disassembly~\cite{yang2026remote} of modular polyhedral architectures. However due to printbed area constraints, traditional printing can typically produce larger overall versions of a model. To address this, future work may explore automatic seam placement for multi-sheet assemblies of objects that exceed printbed size. \martin{Relatedly, \textit{Dapper}~\cite{chen2015dapper} and \textit{Chopper}~\cite{luo2012chopper} partition an input mesh into disjoint parts small enough to be 3D-printed before assembly; however, this voids our method’s benefit of using a single sheet to constrain degrees of freedom via kinematics to simplify assembly.} While our objects are sufficiently mechanically rigid for handling and display, future work could investigate making these demonstrably load-bearing as well. 






\vspace{-1em}
\section{Conclusion}

\system{} introduces a design tool that converts
3D models into flat FDM-printable nets that fold into
shape-accurate 3D objects via living hinges. For complex 3D models, our maximum inscribed rectangular prism approach enables material and time savings on commodity FDM printers while preserving the original mesh surface. For convex polyhedral shapes, a spanning-tree unfolding produces a flat net whose faces exactly match the
original polyhedral geometry, with mass savings but longer printing times. A custom hinge and peg-and-hole insertion pipeline is developed for each method to allow physical folding. We demonstrated the approach on a range of models and evaluated material and time savings, shape accuracy, and repeatability of folding. \system{} represents a step toward the broader goal of shape-accurate, sustainable rapid prototyping on inexpensive desktop hardware.



\bibliographystyle{ACM-Reference-Format}
\bibliography{references}

\appendix

 
\begin{figure*}[h]
  \centering
  \includegraphics[width=0.99\linewidth]
{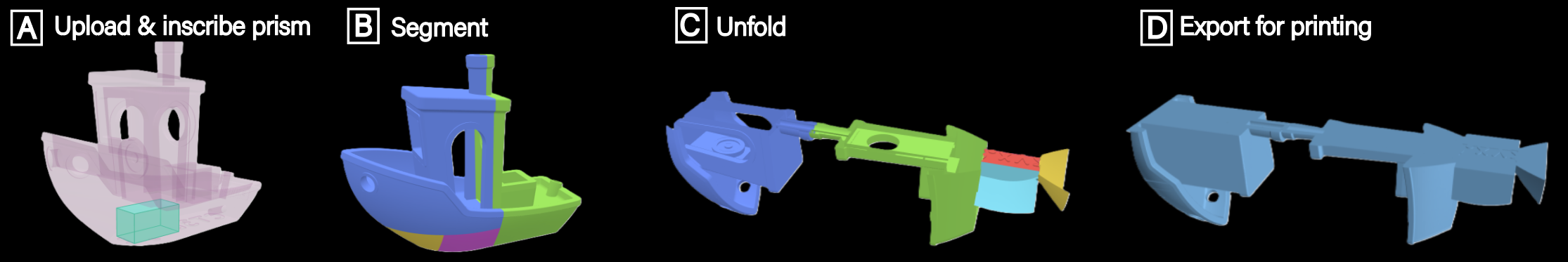}
  \caption{\martin{Unfolding a Benchy: An example of a model with poor volume efficiency ratio. (A) Uploading the model and inscribing the hollow core reveals that only a small internal hollow (turquoise) can be inscribed. This results in a poor mass and time savings, although it is still possible to (B) segment, (C) unfold, and (D) print the model.}}
  \Description{Benchy}
  \label{fig:Benchy}
\end{figure*}

\begin{figure*}[t]
  \centering
  \includegraphics[width=0.99\linewidth]
{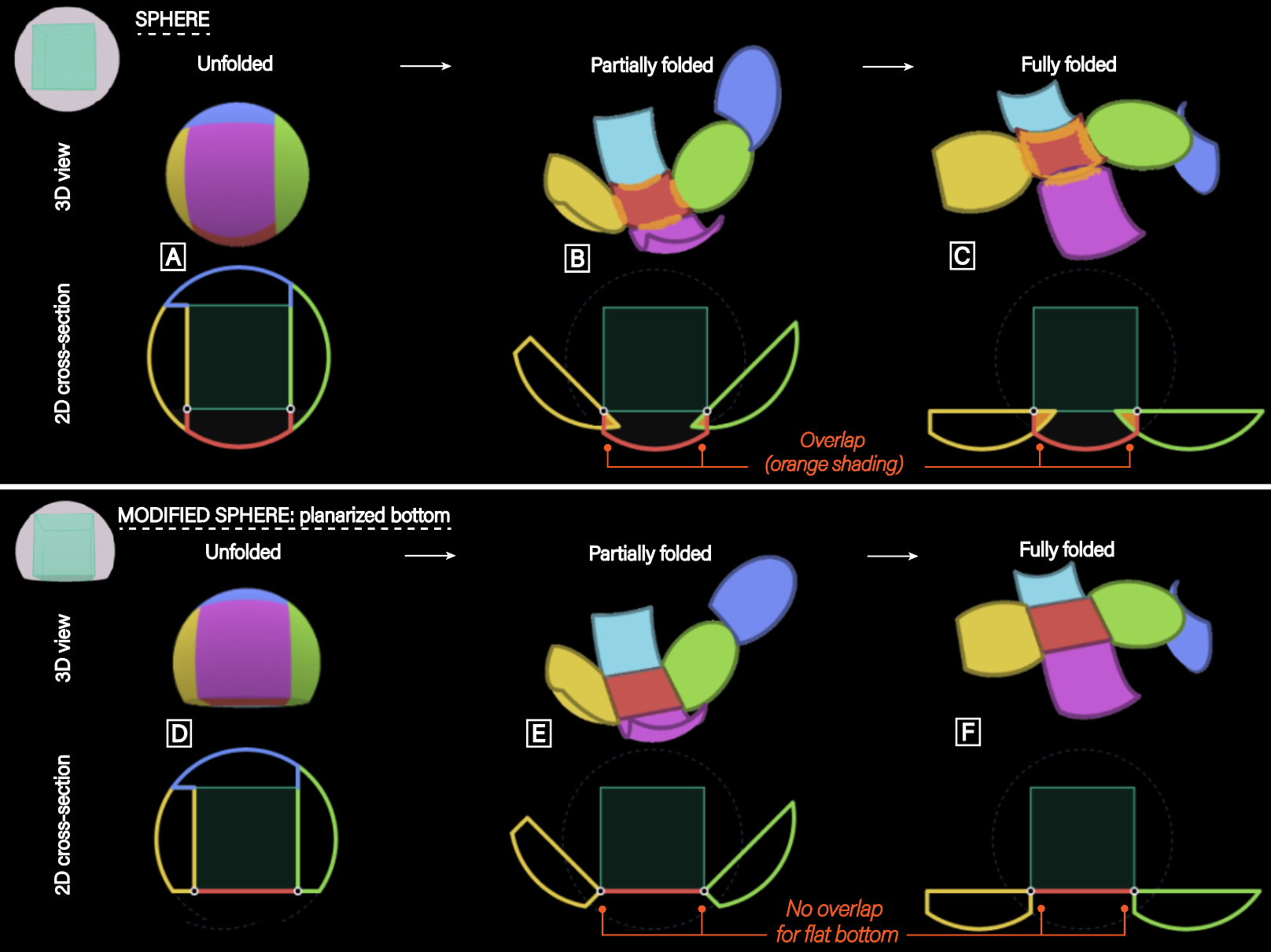}
  \caption{\martin{Visualizing the 1 planar face requirement for overlap-free folding, comparing a sphere (A-B-C) to a modified sphere with a planarized bottom (D-E-F). Both examples depict a full 3D view and a 2D cross-sectional view (rows) in unfolded, partially- and fully-folded keyframes (columns). SPHERE: an inscribed cube segments the sphere into panels (A). Folding the yellow and green side panels around the cube's bottom edges causes both panels to immediately overlap the bottom red panel (B,C). MODIFIED SPHERE: the same inscribed cube segments the modified sphere into panels, but the bottom faces are now all co-planar (D). Folding the yellow and green side panels proceeds without overlapping the bottom red panel (E,F).}}
  \Description{flat-face}
  \label{fig:flat-face}
\end{figure*}

\section{\martin{Unfolding Convex Polyhedra}} \label{Appendix:convex}

The steps taken to unfold and print convex polyhedra are given below. We treat polyhedra as collections of polygons connected at shared hinge edges. The boundary of each panel is the ordered loop of edges that appear exactly once within the panel's triangle set; the dual graph is then built by matching these boundary edges across panels to establish neighboring relationships.




\textbf{Step 1: Spanning-tree traversal:}
Let $\mathcal{F} = \{F_1, \ldots, F_m\}$ be the set of panels in the polyhedron. A face adjacency graph $G = (\mathcal{F}, E)$ is constructed
where edge $(F_i, F_j) \in E$ iff $F_i$ and $F_j$ share at least one
boundary edge. A breadth-first spanning tree $T$ is
rooted at the panel of largest area and
the remaining panels are ordered by BFS discovery order.
Each panel $F_j$ with parent $F_i$ in $T$ is unfolded by the
rotation $R_j = R(\ell_{ij},\, -\theta_{ij})$, where $\ell_{ij}$ is the
shared boundary edge and $\theta_{ij}$ is the signed dihedral angle between the outward normals of $F_i$ and $F_j$. The rotation is applied in the cumulative world frame, so $F_j$ is transformed by
$\prod_{k \in \mathrm{path}(root \to j)} R_k$, a composition of rigid
rotations that exactly undoes the dihedral angle at each hinge. 



\textbf{Step 2: Prepare for 3D printing} Panels are thickened to 3mm and closed to form a printable mesh. At every spanning-tree edge, we 1) generate 0.2mm thick living hinges spanning 95\% of edge lengths, 2) chamfer edges to meet at correct dihedral angles when folded, and 3) generate peg-and-hole connectors on chamferred faces (pegs: 3mm width, 2mm depth, 40\% edge length; holes toleranced to +0.2mm). To mitigate odometrically accumulated fold errors, peg-and-hole connectors are also generated on non-hinge chamferred edges of leaf faces of the spanning tree (holes: 3.5mm width, 2mm depth, 7.5mm length, pegs toleranced to -0.2mm).

\end{document}